\documentclass[sigconf, 9pt]{acmart}

\usepackage{epsfig}
\usepackage{algpseudocode}
\usepackage{algorithm}
\usepackage{subfigure}
\usepackage{hyperref}
\usepackage{flushend}

\setcopyright{acmlicensed}
\begin{document}

\copyrightyear{2018} 
\acmYear{2018} 
\setcopyright{acmlicensed}
\acmConference[IoT S\&P'18]{ACM SIGCOMM 2018 Workshop on IoT Security and Privacy }{August 20, 2018}{Budapest, Hungary}
\acmBooktitle{IoT S\&P'18: ACM SIGCOMM 2018 Workshop on IoT Security and Privacy, August 20, 2018, Budapest, Hungary}
\acmPrice{15.00}
\acmDOI{10.1145/3229565.3229567}
\acmISBN{978-1-4503-5905-4/18/08}



\title[A Library for IoT Privacy-Preserving Traffic Obfuscation]{A Developer-Friendly Library for Smart Home IoT Privacy-Preserving Traffic Obfuscation}

\author{Trisha Datta}
%\orcid{1234-5678-9012}
\affiliation{%
  \institution{Princeton University}
  %\streetaddress{P.O. Box 1212}
  \city{Princeton}
  \state{New Jersey}
  %\postcode{08544}
}
\email{tdatta@princeton.edu}

\author{Noah Apthorpe}
\affiliation{%
  \institution{Princeton University}
  \city{Princeton}
  \state{New Jersey}
}
\email{apthorpe@cs.princeton.edu}

\author{Nick Feamster}
\affiliation{%
  \institution{Princeton University}
  \city{Princeton}
  \state{New Jersey}
}
\email{feamster@cs.princeton.edu}

% The default list of authors is too long for headers.
\renewcommand{\shortauthors}{T. Datta et al.}

\begin{abstract}
The number and variety of Internet-connected devices have grown enormously in the past few years, presenting new challenges to security and privacy. Research has shown that network adversaries can use traffic rate metadata from consumer IoT devices to infer sensitive user activities. 
Shaping traffic flows to fit distributions independent of user activities can protect privacy, but this approach has seen little adoption due to required developer effort and overhead bandwidth costs. 
Here, we present a Python library for IoT developers to easily integrate privacy-preserving traffic shaping into their products. 
The library replaces standard networking functions with versions that automatically obfuscate device traffic patterns
through a combination of payload padding, fragmentation, and randomized cover traffic. 
Our library successfully preserves user privacy and requires approximately 4 KB/s overhead bandwidth for IoT devices with low send rates or high latency tolerances. This overhead is reasonable given normal Internet speeds in American homes and is an improvement on the bandwidth requirements of existing solutions.
\end{abstract}

\settopmatter{printfolios=true}

\maketitle

\section{Introduction}
Over the past few years, the devices that comprise the Internet of Things have been growing in number and variety. According to Business Insider, there will be 24 billion IoT devices in use by 2020 \cite{bus:iot}. While the potential benefits of these devices are well documented, they can also adversely impact personal security and privacy. 

IoT devices often perform very specific functions, making them vulnerable to attackers seeking to infer details about potentially private user behaviors. Apthorpe et al.~\cite{apthorpe:spying} showed that an adversary with access to IoT device traffic flows can detect when user events occur, even when all traffic is encrypted, simply by looking for peaks in traffic send and receive rates. User events differ across IoT devices and can reveal specific information about user activities. A user event for a sleep monitor, for example, usually indicates that a user has fallen asleep or woken up. If adversaries know device identities, they can deduce information about what users are doing inside their homes.

Past research has indicated that traffic shaping by independent link padding (ILP) can protect user privacy from this activity inference privacy risk \cite{apthorpe:spying}. If a traffic flow is shaped so that its send and receive rates (packet send times and sizes) are completely independent of user events, then these rates necessarily leak no information about user behavior. 
Proposed router-based traffic shaping implementations \cite{apthorpe:spying} can protect all IoT devices in the home from an ISP or other WAN observer without requiring device developer cooperation. 
However, router-based solutions do not protect against WiFi observers, such as next-door neighbors or other malicious devices in the home, and require technical knowledge by everyday consumers to set up and manage. 

In this paper, we 
present a Python library that IoT device developers can use to incorporate ILP traffic shaping directly into their devices with minimal effort and for minimal cost. 
The library replaces Python's existing socket library and send()/recv() functions with new versions that automatically obfuscate device traffic patterns
through a combination of packet buffering, payload padding and fragmentation, and randomized cover traffic. 
Developers can call our library's functions to send information between their devices and cloud servers. Our send() function manipulates packet sizes and interpacket intervals, drawing their values from predetermined probability distributions.  
Our receive() function discards cover traffic and recombines fragmented messages.
Our library's familiar interface makes it easy for developers to incorporate into existing code. Because the library must be used by both the sender and receiver, we imagine that the library will be the most useful for devices that communicate only with servers controlled by the device manufacturer.

Our library is currently tailored to 2 different categories of devices based on network behavior with fine-tuning parameters that developers can adjust. 
The first category includes devices that tolerate long latencies in cloud server communications without sacrificing usability. 
The second category includes devices that produce little network traffic overall.
These categories of devices are the most amenable to low-cost ILP traffic shaping and cover a broad range of IoT products. 
We evaluate the overhead cost of traffic shaping with our library when applied to traffic traces from real consumer IoT devices. 
We find that 
use of the library results in approximately 4 KB/s of overhead bandwidth per device, which is acceptable given US internet speeds.

Our library also provides several improvements over router-based traffic shaping. First, on-device shaping does not rely on the ability to install a custom solution on a home router. Second, it provides developers with tools to customize user data obfuscation for their device characteristics instead of providing a generic solution at the router level that is applied to all devices. Third, our library prevents long delays that could affect device functionality. Fourth, while traffic shaping performed at the router does not protect against an attacker who is sniffing wireless traffic in the home, our library protects against this threat by obfuscating traffic before it leaves the device.

\begin{sloppypar}
The library is available at \url{https://github.com/TrishaDatta/PrivacyPreservingTrafficObfuscation}~\cite{so:py}.
\end{sloppypar}

\section{Related Work}

Our primary motivation is to prevent attackers from using encrypted IoT device traffic to infer user events that would violate the privacy of homeowners. Apthorpe et al.~recently showed that by looking for peaks in traffic flow patterns, attackers can detect user events for many different kinds of IoT devices, even when traffic is encrypted~\cite{apthorpe:spying}. Their threat model consisted of an attacker with abilities similar to an ISP, who can observe Internet traffic in and out of a smart home. The attacker's goal is to identify IoT devices inside the home and infer user behaviors from traffic rates. 
We assume the same threat model in this paper.

Previous work has explored how to protect user anonymity from traffic flow analysis. Most solutions in this space rely on either independent link padding (ILP) or dependent link padding (DLP) \cite{dyer:peek, fu:analytical, sham:timing, wang:dep}. ILP forces traffic to fit a predetermined distribution by padding and fragmenting packets and sending cover packets. DLP also uses padding, fragmentation, and cover traffic but uses some information from the real traffic flow to set the schedule for packet sending. Forcing traffic to fit pre-specified distributions is one of the primary tools we use in our solution.

The two main parameters of ILP and DLP are packet sizes and interpacket delays. Past work has explored how these two variables can best be altered to protect user privacy. Dyer et al.~have shown that simply padding or fragmenting packet lengths to a constant size is not sufficient for hiding user activity \cite{dyer:peek}, and Fu et al.~have shown that variable interpacket delays obfuscate user activity better than constant interpacket delays \cite{fu:analytical}. Past research has also shown that DLP can preserve anonymity in systems susceptible to traffic flow analysis attacks \cite{sham:timing, wang:dep}. However, as pointed out by Apthorpe et al.~\cite{apthorpe:spying}, because these DLP algorithms preserve the relative relationships between time intervals with high rates of traffic and time intervals with low rates of traffic, the attack we are concerned with would still be feasible. 

\section{Threat Model \& Target Devices}
\label{techapp}

Our threat model assumes that the adversary knows which IoT devices are in a house, the identity (e.g., MAC addresses) of the IoT device,  and the sizes and times of the packets each device sends. We assume that all packet payloads are encrypted. The adversary's goal is to determine when user events occur. This goal is intentionally broad in order to encompass many different kinds of devices. For example, a user event for a sleep monitor might be 
someone falling asleep or waking up.

We split smart home devices into three categories based on their network behavior:
\begin{enumerate}
    \itemsep0em 
    \item High-latency devices
    \item Low-latency low-bandwidth devices
    \item Low-latency high-bandwidth devices
\end{enumerate} 
The functionality of ``high-latency'' devices is not affected by longer network delays, while ``low-latency''  devices require fast round trip times for acceptable performance. For example, a sleep monitor is a high-latency device because its functionality does not rely on quick server responses. Conversely, a light switch that can be turned on and off by an app is a low-latency device; the user expects minimal delay between app interaction and the bulb turning on or off. 

We further divide low-latency devices into high-bandwidth and low-bandwidth categories based on the amount of traffic they send during user events.
Note that we have not defined a hard line between categories. Developers who feel that their devices may straddle two categories may want to try our solution for each category and see which works best for their devices.

In this paper, we focus only on high-latency devices and low-latency low-bandwidth devices. 
These categories include many smart home devices 
that perform specialized functions,
such as sleep monitors, embedded sensors, motion detectors, lightbulbs, and kitchen appliances. 
The user activity inference attack described in~\cite{apthorpe:spying}
is particularly harmful for special purpose IoT devices.  Network traffic peaks from these devices imply specific user activities without needing additional information. 
By creating a traffic shaping library for these device categories, we can improve privacy for a wide range of IoT devices.

Additionally, traffic from high-latency devices and low-latency low-bandwidth devices is particularly amenable to traffic shaping. These devices can either tolerate imposed delays or require less cover traffic to mask user activities, respectively.

In contrast, the traffic from low-latency high-bandwidth devices is difficult to shape without impacting useability or imposing excessive overhead. 
Obfuscating user activities in traffic from these devices requires either large amounts of cover traffic or a mechanism to produce ``fake'' user events. The former is not ideal, especially for users with data caps; the latter is challenging and the subject of ongoing work outside the scope of this paper.

\begin{figure}[t]
\centering
\epsfig{file=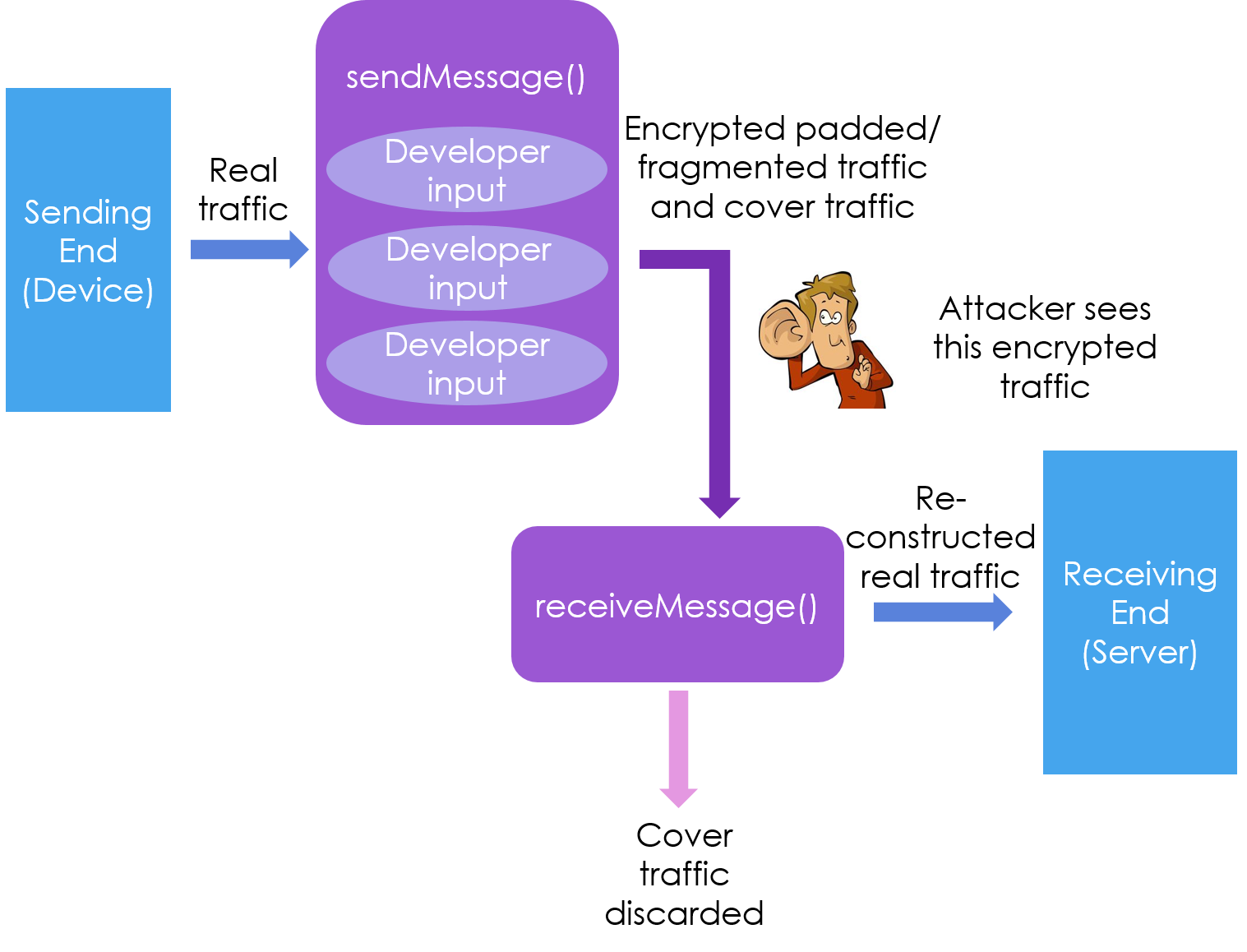,width=8cm}
\caption{Overview of traffic obfuscation library. }
\label{fig:lib}
\end{figure}

\section{Traffic Shaping Library}
Our library consists of a set of networking primitives that shape traffic to distributions independent of user activities (Figure~\ref{fig:lib}). This prevents attackers from inferring user activities from network traffic patterns. 

Our library performs traffic shaping in the sending direction by padding payloads, fragmenting payloads, and adding cover packets. 
Because we assume that traffic is encrypted,  cover traffic consists of random bytes.
The send function adds 6-7 bytes of overhead (a recovery header) indicating fragmentation and payload padding details. 
This allows the receive function to discard cover traffic and recover messages in their original form. 

We implemented our library in Python because it is often used in IoT development. The application-facing behavior of our library's send/receive functions are as similar to their standard Python counterparts as possible. This  will facilitate easier seamless integration of our library into actual code. Users of our library can also provide inputs to control padding and fragmentation parameters.

Our library allows developers flexibility when programming their devices and does not interfere with device functionality or user experience. 
It also gives developers a easy way to obfuscate user data without relying on router-based or other in-network protections. This places the burden of privacy protection on the device developers rather than smart home owners.

\subsection{Sender API Description}
The sending API of our library is implemented as a \texttt{Sender} object with the following methods:

\begin{enumerate}
    \item 
\texttt{Sender.Sender(host, port, $D$, $X$)} creates a TCP socket between the sender and the receiver specified by \texttt{host} and \texttt{port}. $D$ and $X$ specify parameters for the interpacket interval and payload size distributions respectively. We also initialize a message queue $q$.

\item \texttt{Sender.send(msg)} adds \texttt{msg} to the message queue. 

\item \texttt{Sender.startPeriodicallySending()} starts the sending of cover traffic. It should be called once before any calls to \texttt{Sender.send(msg)}. It starts a thread that implements Algorithm~\ref{alg1}.

\begin{algorithm}[t]
\begin{algorithmic}
\While{doSend}
\State $d \Leftarrow$ Sample from distribution $D$
\State Pause for $d$ seconds
\State $x \Leftarrow$ Sample from distribution $X$
\If{$q$ is not empty}
  \State{$m \Leftarrow$ first message in $q$}
  \State{$len \Leftarrow$ size($m$)}
  \If{$len \le x$}
    \State Pad $m$ to $x$ bytes.
  \Else
    \State Put last $len - x$ bytes of $m$ at head of $q$
    \State $m \Leftarrow$ first $x$ bytes of $m$
  \EndIf
  
\Else
  \State{$m \Leftarrow$ Random cover traffic of length $x$}
\EndIf
  \State $m \Leftarrow$ $m + $ recovery header
  \State Send $m$
\EndWhile
\end{algorithmic}
\caption{ILP Traffic Shaping}
\label{alg1}
\end{algorithm}

 \item \texttt{Sender.close()} closes the socket. Because there might still be messages in the queue, it waits until the queue is empty (i.e., the separate thread has finished sending all messages in the queue).

\end{enumerate}

Because high-latency devices are more tolerant of variable delays, our high-latency solution uses random distributions for $D$ and $X$. For low-bandwidth low-latency devices, $D$ and $X$ are constants; this enforces a constant sending rate. The developer must ensure that the parameters $D$ and $X$ allow a send rate that is higher than the rate at which the device would normally send data to prevent functionality from being affected. There is thus a trade-off between latency and overhead bandwidth.

\subsection{Receiver API Description}

Our receiving library is much simpler than our sending library because the receiving side just reconstructs the original messages using the recovery headers. There is only one function on the receiving side: \texttt{recv(s, conn)}. It takes in the socket \texttt{s} and the connection \texttt{conn} returned by \texttt{s.accept()} and returns the original messages sent by the sending side. This function replaces \texttt{conn.recv()}.

\section{Performance Evaluation}
We evaluate our traffic shaping library using two criteria: privacy protection and overhead bandwidth. 

\subsection{Data Collection}
We test the library with traffic collected from real IoT devices. The traffic was collected as part of the work described in \cite{apthorpe:spying}. 
Ideally, testing would involve deploying our library on multiple real IoT devices. As an approximation, we use the collected traffic to simulate IoT device traffic using our library. 
The simulation involves performing the following operations on each packet recorded from each device:
\begin{enumerate}
    \itemsep0em 
    \item Extract the payload length and send time of each packet. 
    \item Replay the traffic between two laptops on the same WiFi network using our library's send and receive functions. 
    \item Record the resulting traffic between the laptops using WireShark.
    \item Repeat this procedure, experimenting with a range of parameters for each device to measure parameter effects on bandwidth consumption.
\end{enumerate}

\begin{figure}[t]
\centering
\epsfig{file=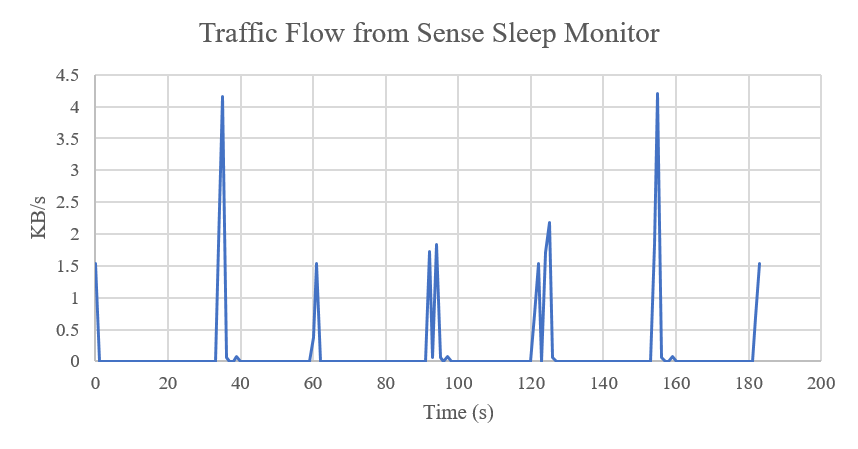,width=8cm}
\epsfig{file=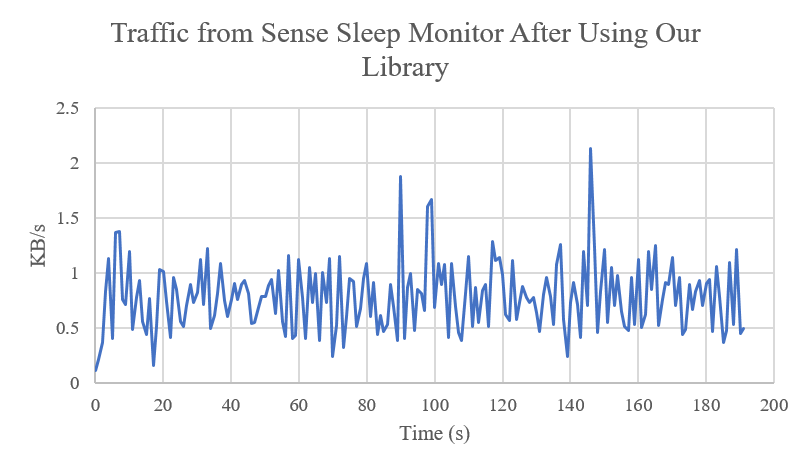,width=8cm}
\caption{\textit{Top:} Traffic flow from a Sense Sleep Monitor when our library is not used. \textit{Bottom:} Traffic flow from a Sense Sleep Monitor with the effects of our library. Payload sizes are drawn from a uniform distribution of $[50, 200]$ bytes and interpacket delays from a uniform distribution of $[0, 0.6]$ seconds.}
\label{fig:sense2}
\end{figure}

\begin{figure}[t]
\epsfig{file=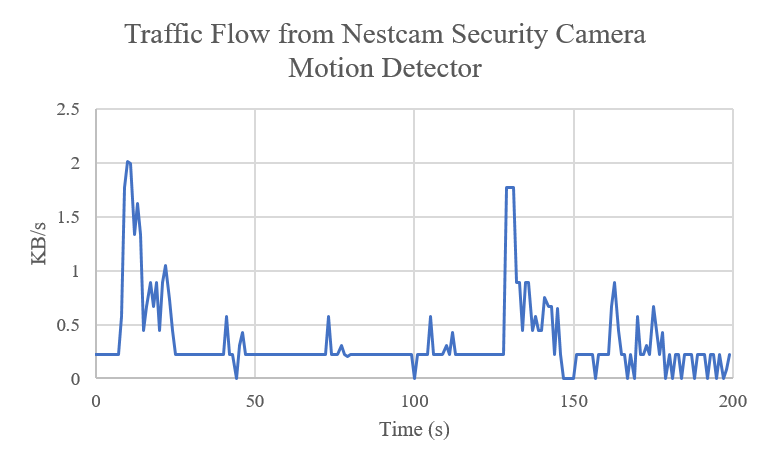,width=8cm}
\epsfig{file=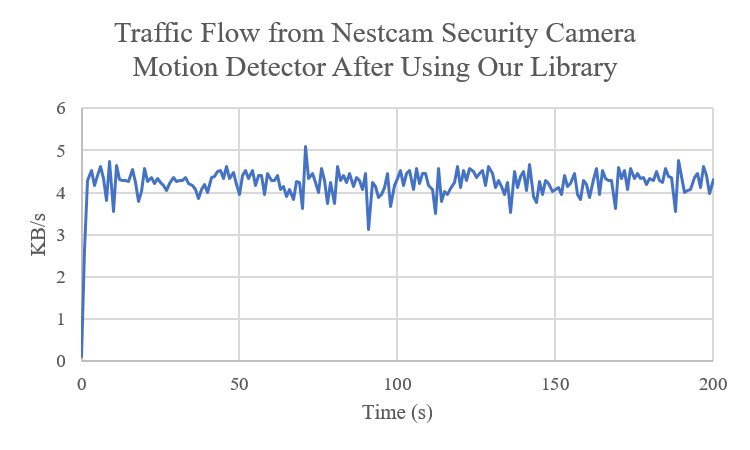,width=8cm}
\caption{\textit{Top:} Traffic flow from a Nestcam Security Camera motion detector when our library is not used. \textit{Bottom:} Traffic flow from a Nestcam Security Camera motion detector with the effects of our library. The constant interpacket delay is $0.05$ seconds, and the constant payload size is $120$ bytes.}
\label{fig:nest2}
\end{figure}

\subsection{Privacy Preservation Evaluation}
Our library completely obfuscates user activity because payload sizes and interpacket intervals are drawn from predetermined distributions independent of user activities.  
These features therefore do not contain information that would allow an adversary to infer user activities.
Figures~\ref{fig:sense2}~and~\ref{fig:nest2} show traffic from a Sense Sleep Monitor, a high-latency device, and a Nestcam Security Camera in low-latency low-bandwidth motion-detection mode, respectively, with and without our library.
Clearly, the shaped traffic does not preserve user event peaks.

\begin{figure}[tp]
\centering
\begin{subfigure}{}
\epsfig{file=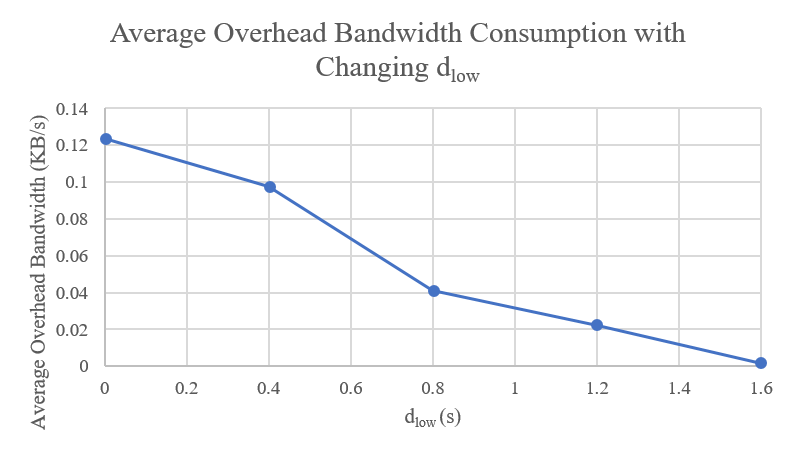,width=8cm}
\end{subfigure}\par\medskip
\begin{subfigure}{}
\epsfig{file=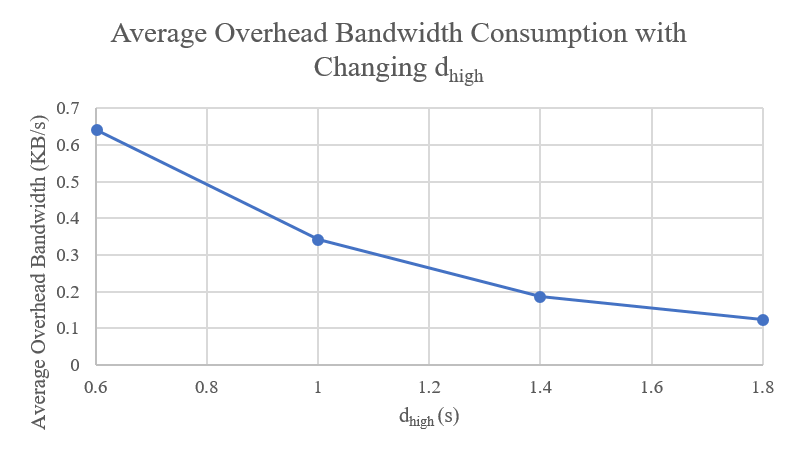,width=8cm}
\caption{The effects of changing $d_{low}$ (\textit{top}) and $d_{high}$ (\textit{bottom}) on additional bandwidth consumed for a Sense Sleep Monitor, a high-latency device.}
\label{fig:dhigh}
\end{subfigure}\par\medskip
\end{figure}

\begin{figure}[tp]
\centering
\epsfig{file=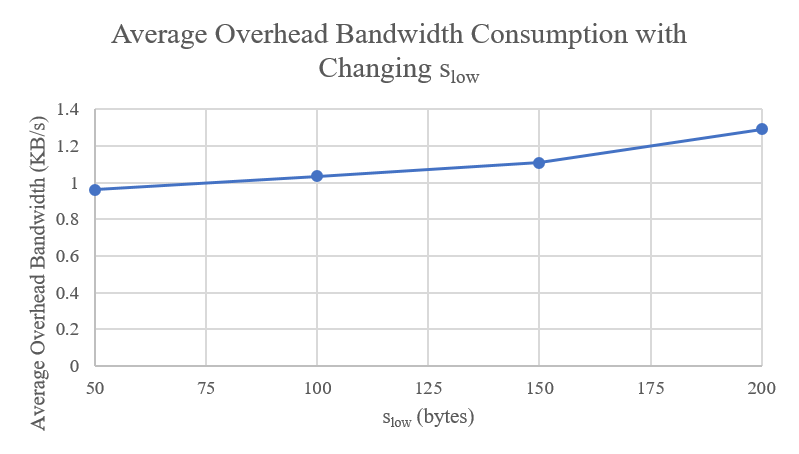,width=8cm}\\
\epsfig{file=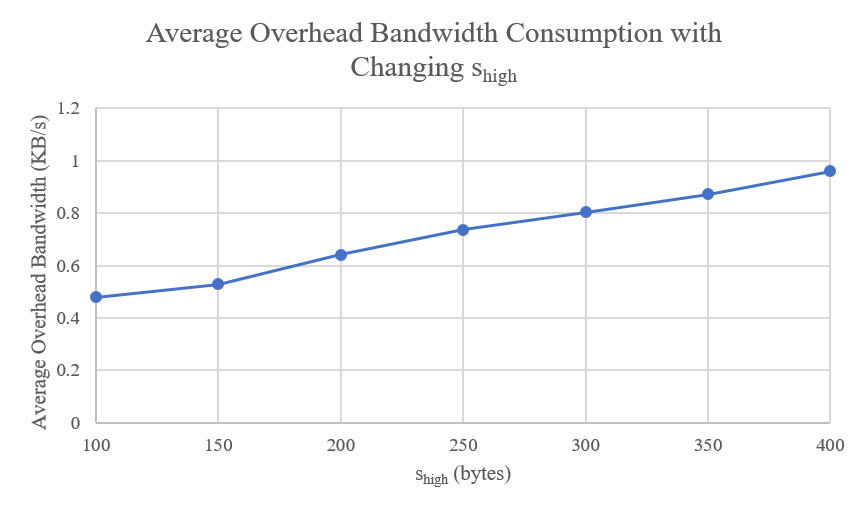,width=8cm}\\
\epsfig{file=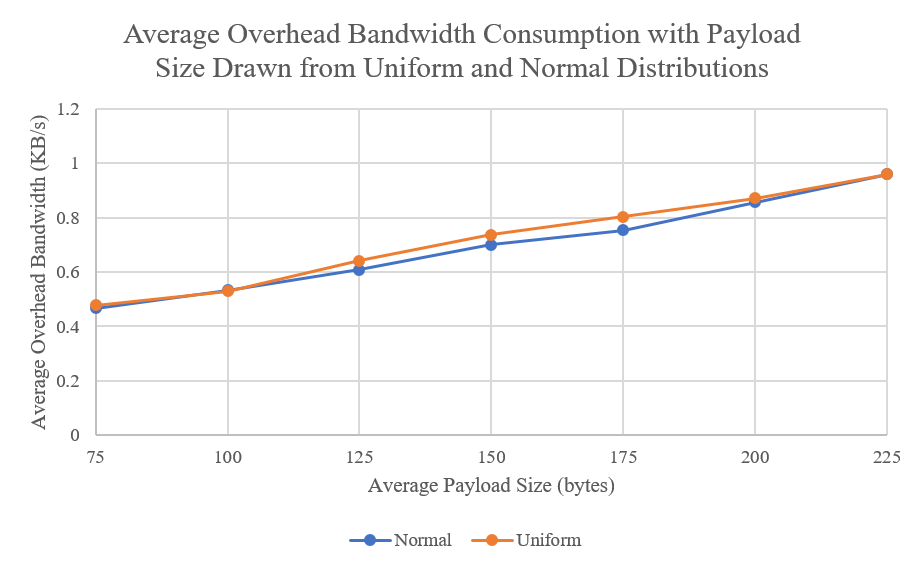,width=8cm}
\caption{The effects of changing $s_{low}$ (\textit{top}) and $s_{high}$ (\textit{middle}) on additional bandwidth consumed for a Sense Sleep Monitor, a high-latency device. \textit{Bottom:} The effect of using a normal distribution vs.~a uniform distribution for the payload size distribution for a Sense Sleep Monitor, a high-latency device.  }
\label{fig:normuni}
\end{figure}

\subsection{Bandwidth Consumption Evaluation}
In this section, we examine how changing interpacket delay and packet size distributions parameters affects overhead bandwidth consumption.

\subsubsection{High-Latency Device}
The average rate of bandwidth consumption for $200$ seconds of traffic from a Sense Sleep Monitor without the effects of our library is $143.78$ bytes/second. We test our library using a uniform interpacket delay distribution defined by the interval $[d_{low}, d_{high}]$. 
Increasing either $d_{low}$ or $d_{high}$ decreases the overhead bandwidth (Figure \ref{fig:dhigh}). This makes sense because we are decreasing the rate at which packets are being sent. Increasing average interpacket delay can increase the total time it takes to send out all of the packets. However, because the sleep monitor is a high-latency device, this is not a major concern. 

\begin{figure}[t]
\centering
\begin{subfigure}{}
\epsfig{file=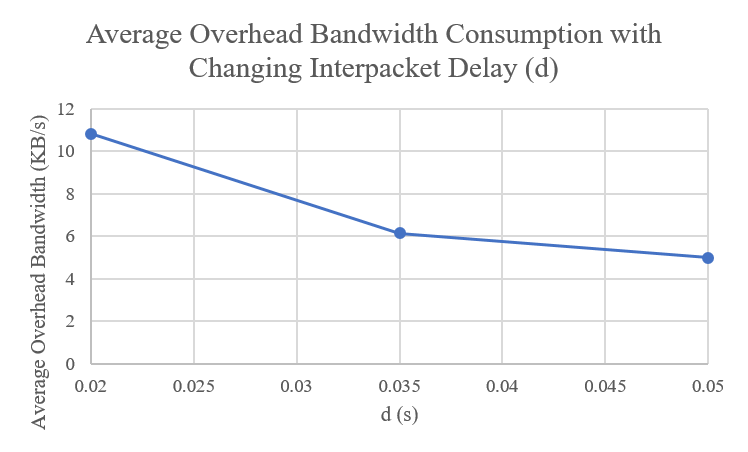,width=8cm}
\end{subfigure}\par\medskip
\begin{subfigure}{}
\epsfig{file=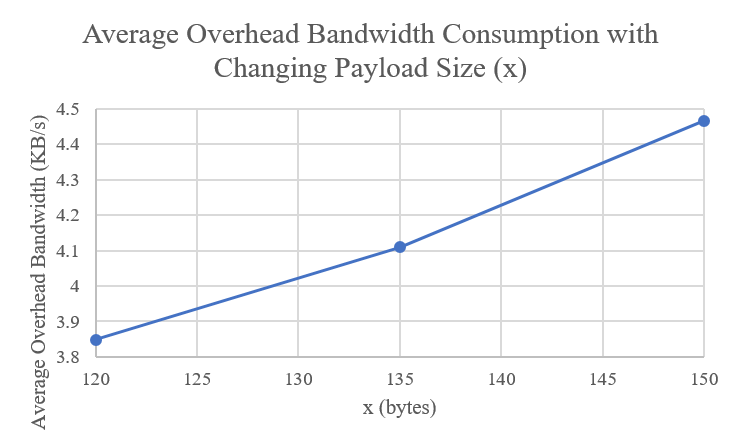,width=8cm}
\caption{The effects of changing the interpacket delay (\textit{top}) and payload size (\textit{bottom}) on additional bandwidth consumed for a Nestcam Security Camera Motion Detector, a low-bandwidth low-latency device. }
\label{fig:ipdpkt}
\end{subfigure}\par\medskip
\end{figure}

We next test the effect of payload size distribution. We use a uniform distribution defined by $[s_{low}, s_{high}]$.
Increasing either $s_{low}$ or $s_{high}$ causes the amount of additional bandwidth required to increase (Figure \ref{fig:normuni}). Decreasing either $s_{low}$ or $s_{high}$ can increase the total time it takes to send all of the packets. This delay increases as the average payload size decreases and can add several seconds to the overall time. However, because we are dealing with high-latency devices, this is not a major concern. We also look at using a uniform distribution versus a normal distribution for packet sizes (Figure \ref{fig:normuni}). We find that the average overhead bandwidth consumption is very similar.

The range of parameters discussed in this section uses up to approximately 1.5 KB/s in overhead bandwidth.

\subsubsection{Low-Bandwidth Low-Latency Device}
Developers can tune two parameters when using our library for low-bandwidth low-latency devices: the constant interpacket delay $d$ and the constant payload size $x$. We examine the effect of changing both of these parameters on overhead bandwidth consumption for $200$ seconds of traffic from a Nestcam Security camera in motion detection mode. The average bandwidth of this traffic without our library is $346.04$ bytes/second. As could be expected, increasing the interpacket delay decreases the average additional bandwidth, and increasing the payload size increases the average additional bandwidth (Figure~\ref{fig:ipdpkt}). The range of parameters discussed in this section uses approximately 3.8 to 11 KB/s overhead bandwidth.

\subsubsection{Discussion: Overhead Bandwidth Costs}
In general, we can find parameter settings that consume around 4 KB/s of overhead bandwidth per device. This is a substantial improvement upon the solution presented in \cite{apthorpe:spying}, which used 40 KB/s of overhead bandwidth for three devices. A device that requires 4 KB/s of overhead bandwidth will require  around 10.4 GB of overhead bandwidth a month. This overhead is reasonable in the context of an American smart home. Comcast and Cox Communications, the largest and third largest cable internet providers in the US by number of subscribers \cite{broad:prov}, both have data caps of 1 TB/month and charge an additional \$10 for every additional 50 GB used \cite{ars:tech}. The 10.4 GB/month per IoT device would only consume around 1\% of a 1 TB data cap.

\section{Conclusion}
The privacy and security issues involved with IoT devices will only continue to proliferate as the number of IoT devices grows. In this work, we present a Python library of networking primitives that smart home device developers can use to help protect their users' privacy from adversaries attempting to infer user activities from encrypted traffic.
Our library uses random interpacket delays, fragmented/padded messages, and cover traffic to force traffic to fit predetermined distributions. 
The library is targeted towards smart home devices with high latency tolerance or low bandwidth usage. 

We evaluate our solutions on two criteria: privacy preservation and additional bandwidth consumption. Our library preserves user privacy because the resultant traffic flows are completely independent of user activities.  Our library also uses less than 4 KB/s of overhead bandwidth for a wide range of parameters. This overhead data usage is reasonable given the rate limits of typical U.S. homes and is a substantial improvement on previous solutions. 

\bibliographystyle{ACM-Reference-Format}
\bibliography{IoT-traffic-obfuscation}

\end{document}